\documentclass[conference,a4paper]{IEEEtran}
\usepackage{cite}

\ifCLASSINFOpdf
  \usepackage[pdftex]{graphicx}
  \graphicspath{{../pdf/}{../jpeg/}}
  \DeclareGraphicsExtensions{.pdf,.jpeg,.png}
\else
\usepackage[dvips]{graphicx}
\graphicspath{{../eps/}}
\DeclareGraphicsExtensions{.eps}
\fi
\usepackage{listings}
\usepackage{multirow}
\begin{document}
%
\title{Implementation and evaluation of data-compression algorithms
  for irregular-grid iterative methods on the PEZY-SC processor}


%

%
\newcommand{\nonbdmode}{1}
\ifnum \nonbdmode=1

\author{\IEEEauthorblockN{Naoki Yoshifuji\IEEEauthorrefmark{1},
Ryo Sakamoto\IEEEauthorrefmark{2},
Keigo Nitadori\IEEEauthorrefmark{3}and 
Jun Makino\IEEEauthorrefmark{4}}
  \IEEEauthorblockA{\IEEEauthorrefmark{1}
Fixstars,  18F GateCity Osaki West Tower,  1-11-1 Osaki, Shinagawa-ku,
     Tokyo 141-0032,  Japan\\
   Email: yoshifuji@fixstars.com}
\IEEEauthorblockA{\IEEEauthorrefmark{2}Fixstars,  18F GateCity Osaki West Tower,  1-11-1 Osaki, Shinagawa-ku,
     Tokyo 141-0032,  Japan\\
   Email: sakamoto@fixstars.com \\
{\footnotesize Present Address: PEZY Computing, 5F Chiyoda Ogawamachi Crosta, 1-11 Kanda Ogawamachi Chiyoda-ku, 101-0052 Tokyo, Japan}\\
   {\footnotesize Present Email: sakamoto@pezy.co.jp}}
\IEEEauthorblockA{\IEEEauthorrefmark{3}RIKEN AICS, 7-1-26 Minatojima-minami-machi, Chuo-ku, Kobe,
  Hyogo 650-0047, Japan\\
  Email: keigo@riken.jp}
\IEEEauthorblockA{\IEEEauthorrefmark{4}Department of Planetology, 
   Kobe University,
  1-1 Rokko-dai, Nada-ku,
   Kobe 657-8501,  Japan\\
   Email: jmakino@people.kobe-u.ac.jp}}

\fi


\maketitle

\begin{abstract}
  Iterative methods on irregular grids have been used widely in all
  areas of comptational science and engineering for solving partial
  differential equations with complex geometry. They provide the
  flexibility to express complex shapes with relatively low
  computational cost. However, the direction of the evolution of
  high-performance processors in the last two decades have caused
  serious degradation of the computational efficiency of iterative
  methods on irregular grids, because of relatively low memory
  bandwidth. Data compression can in principle reduce the necessary
  memory memory bandwidth of iterative methods and thus improve the
  efficiency. We have implemented several data compression algorithms
  on the PEZY-SC processor, using the matrix generated for the HPCG
  benchmark as an example. For the SpMV (Sparse Matrix-Vector
  multiplication) part of the HPCG benchmark, the best implementation
  without data compression achieved 11.6Gflops/chip, close to the
  theoretical limit due to the memory bandwidth. Our implementation
  with data compression has achieved 32.4Gflops. This is of course
  rather extreme case, since the grid used in HPCG is geometrically
  regular and thus its compression efficiency is very high. However,
  in real applications, it is in many cases possible to make a large
  part of the grid to have regular geometry, in particular when the
  resolution is high. Note that we do not need to change the structure
  of the program, except for the addition of the data
  compression/decompression subroutines.  
  Thus, we believe the data compression will be
  very useful way to improve the performance of many applications
  which rely on the use of irregular grids.
\end{abstract}


%
\IEEEpeerreviewmaketitle


\section{Introduction}
\label{sect:introduction}

In this paper, we describe the implementation and performance of
the multiplication of sparse matrix and vector (hereafter the SpMV
multiplication) on the PEZY-SC
processor. In particular, we focus on the effect of various data
compression schemes on the performance.
The multiplication of sparse matrix and vector is the most time
consuming part of many real applications which use irregular grids.
The best-known example is the FEM (Finite Element Method) for
structural analysis and many other CAE applications.  Irregular grids
are essential to allow the analysis of objects with complex shapes.
Though it is not impossible to apply regular grids to complex shapes,
generally irregular grids offer more accurate results with much
smaller number of freedoms.

However, it has become very difficult to achieve even reasonable
efficiency on the SpMV multiplication on modern HPC systems. There are
two main reasons for this difficulty. The first one is the memory
bandwidth. Consider the multiplication of matrix $A$ and vector $x$,
\begin{equation}
  y = Ax.
  \label{eq:SpMV}
\end{equation}
For real applications, the matrix $A$ is too large to fit to the cache
memory. On the other hand, vectors $x$ and $y$ are much smaller, and
there is always the possibility of extensive data reuse for
them. Thus, the dominant part of memory access for an SpMV operation
is the reading of the (sparse) matrix $A$.

The exact data size of the matrix $A$ depends on the used data format, but
it cannot be smaller than the number of non-zero elements of
$A$. The number of floating-point operation per one non-zero element
of $A$ is two. Thus, If the data format is used is the
double-precision format, memory read of eight bytes takes place for
every two floating point operations. In other words, the ``required''
B/F (byte per flops) number is $8/2=4$. Note that here we ignored the
memory read for the indices. If we use the ELL format, which usually
is the most efficient format for storing the matrix in FEM
applications, required bandwidth can increase by 50-100\%. Thus, the
required memory bandwidth, in terms of the B/F number can be between
six and eight.

In the 1980s, vector supercomputers had the memory subsystem which
could support at least a fair fraction of the the memory bandwidth
requirement of the SpMV operation, since the hardware B/F number of
vector machines in 1980s ranged between 4 and 12. Here, the hardware
B/F number of a machine is defined as the theoretical (or measured)
memory bandwidth in bytes/second divided by the theoretical peak
performance of the floating-point operation measured in the number of
floating-point operations per second. Vector machines in 1980s had
sufficient memory bandwidth to keep the floating-point unit busy for
SpMV multiplications.

However, the B/F numbers of microprocessors used for modern HPC
systems are much smaller. For example, the B/F number of K computer is
0.5, and this is rather exceptionally high in today's
standard. Recent Xeon-based systems have the B/F numbers of around
0.2. If the required B/F number is six, this means that the
theoretical maximum efficiency of modern HPC systems would be around or
less than 5\%. 

In fact, the ratio between the measured HPL performance and measured
HPCG performance of machines in the June 2016 top 10 list of HPCG
benchmark\footnote{http://www.hpcg-benchmark.org/custom/index.html?lid=155\&slid=288}
ranges between 0.4 and 5\%, and the numbers of Xeon-based systems are
2-3\%.

Thus, the low memory bandwidth of modern HPC systems is clearly the
primary reason for the very low efficiency of SpMV multiplication on
them.

The second reason is the tendency of designers of modern to adopt SIMD arithmetic unit with
rather wide width (4-8 words). The computing  kernel of SpMV
operation, for irregular matrices, requires the indirect access of the
elements of either the vector or the matrix. The performance of
indirect access on modern processors with wide SIMD units are very
low. Some  processors do not support  SIMD operations for indirect
memory access. Even on machines with SIMD instructions for indirect
memory access, their throughput is much lower than simple SIMD
load/store instructions.  On some machines this inefficiency can cause
further degradation of performance of SpMV operations.

One way to reduce the required memory read is the element-by-element
(EbE) method. In the EbE method, the sparse matrix $A$ is constructed
from the original physical and topological data of each element on the
fly. Since the total amount of data for all elements is significantly
smaller than the size of the generated matrix, we can reduce the
amount of memory access. Even though the calculation cost of
on-the-fly construction of the matrix is fairly high, the total
computing time can be significantly reduced by moving to the EbE
method. EbE method has become widely used in many FEM applications,
since it can achieve quite significant improvement in the actual speed
of calculation, even though the calculation cost is increased.

Another way which can potentially be useful in reducing the required
amount of memory access is to compress the matrix using some data
compression algorithm. However, even though there are many research
papers on the use of data compression in HPC applications, there seems
to be little work on the application of data compression to SpMV
multiplications. One possible reason is that in order to achieve
actual speedup, data decompression algorithm must be extremely
efficient, since the number of floating-point operations per one
matrix element is only two. If the decompression algorithm requires
more than a few instructions, it will cause quite significant increase
of the total cost. Moreover, generally the decompression algorithm
requires some table-lookup operations, in other words, the indirect
memory access for which modern microprocessors are not particularly
efficient.

On the other hand, if the hardware B/F number of a system is extremely
low, we might be able to achieve significant performance improvement
on SpMV multiplications by using data compression/decompression.

In this paper, we report the performance of the ZettaScaler-1.5
supercomputer\cite{Ishii2015} on the SpMV part of HPCG
benchmark\cite{Dongarraetal2015,  Herouxetal2013}, with and without
the use of data compression/decompression.

The ZettaScaler (previously called ExaScaler) system is based on the
first-generation PEZY-SC 1024-core processor chip.
It  appeared in the
TOP500 list of November 2014 and ranked \#2 in the Green500 list.
In the June 2015 Green500 list, three ExaScaler systems occupied top
three ranks. The system listed \#1 achieved the performance per watt
exceeding 7 Gflops/W, significantly higher than the number of the \#1
system for November 2014 Green500 list. As of June 2016, it still
keeps the \#1 position in the Green500 list.

The reason why we used the ZettaScaler system as the testbed for the
data compression algorithm is that its hardware B/F number is rather 
low, around 0.05. Thus, it is ideal as the testbed of algorithms which
will be useful for processors in the near future. 
In addition, its processor cores do not have SIMD
units. Thus, we might be able to achieve pretty good speedup for SpMV
multiplication using data compression.

The ZettaScaler system is rather similar to modern GPGPU-based
systems, in which the GPGPUs are connected to Intel Xeon processors
through PCIe interface, and Xeon processors are connected using
Infiniband network. There are, however, two unique features of the
ZettaScaler system. The first one is of course the use of PEZY-SC
processor chip, which is a 1024-core MIMD processor with physically
shared memory and  hierarchical cache. It  was developed by a Japanese
venture company, PEZY Computing. The second feature is the immersion
cooling system in which fluorocarbon (3M Fluorinert FC-43) is used.
For the ZettaScaler system, they designed  motherboards for Xeon
processors and processor cards for PEZY-SC processors to achieve
high-density packing. The use of immersion cooling has the potential
advantage of reducing the PUE number and also reducing the junction
temperature of the processor chips, resulting in more energy efficient
operation. However, the primary reason of the high
performance-per-watt number of the ZettaScaler system is the design of
the PEZY-SC processor itself.

The PEZY-SC processor integrates 1024 MIMD cores, each with fully
pipelined double-precision multiply-and-add (MAD) unit, into a die of size 400 
mm$^2$,  using TSMC's 28HPM process. Its nominal power consumption is
only 65 W for the operation with 733 MHz clock.

At least for the HPL benchmark, or more specifically the DGEMM operation
(double precision dense matrix multiplication), the PEZY-SC processor
has achieved quite impressive performance per watt, even though the
efficiency compared to the theoretical peak performance is still
rather low (slightly better than 50\% for HPL). On the other hand, the porting
of applications could be relatively easy, since the PEZY-SC processor
is an MIMD manycore processor with hierarchical (but non-coherent)
cache and physically shared memory. Also, a fairly well designed
subset of OpenCL, PZCL, is supported.

In this paper we first present the performance of HPCG on PEZY-SC
processor, with usual optimizations applied in previous works. Then we
proceed to present the performance of ``optimized'' implementations of
SpMV operation  with
on-the-fly data compression and decompression.



This paper is organized as follows. First, in section
\ref{sect:PEZY_SC_description}, we present the overview of the PEZY-SC
processor and the ZettaScaler system.  In section
\ref{sect:HPCG_on_PEZY_SC}, we describe our implementation of HPCG on
PEZY-SC processor.  In section \ref{sect:HPCG_result}, we present the
performance result.  Finally, in section \ref{sect:SpMV_tuning}, we
present data compression/decompression algorithms we implemented 
and its measured performance on PEZY-SC.  In
section \ref{sect:discussion}, we summarize the paper and discuss the future
directions for research and development.

\section{The PEZY-SC processor chip and the ZettaScaler system}
\label{sect:PEZY_SC_description}

In this section, we overview the architecture of the PEZY-SC processor
and the ZettaScaler system.
In subsection \ref{sect:PEZY_SC_processor}, we give a brief overview
of the PEZY-SC processor. 
In subsection \ref{sect:ZettaScaler}, we give a brief overview
of the ZettaScaler system.

\subsection{The PEZY-SC processor}
\label{sect:PEZY_SC_processor}

The PEZY-SC processor\cite{PEZYSCDataSheet,Ishii2015} integrates 1024 MIMD cores with three levels of
cache memory. In this section, we describe the structure bottom-up,
starting from the processor core.

Each of the PEZY-SC core can do one double-precision MAD operation or
two single-precision MAD operations per cycle. It has the usual
load-store architecture. The core is a quite simple dual-issue, in-order
core with four-way (can be eight-way) SMT. Thus, the impact of the
latency of data caches to the performance is relatively small, even
though the in-order core is used.

One rather unusual feature of the PEZY-SC processor core is that each
core has 16KB of the local memory, accessible only by that core. It
has the separate local address space, and can be used to store the data
which is repeatedly used by the core. Since this local memory provides
the largest on-chip storage (16MB in total) with very high bandwidth,
it is essential to take advantage of this local storage to achieve
high efficiency, in particular for compute-intensive applications.
For example, there is well-known tradeoff between the required memory
bandwidth and required on-chip storage, for the performance of the
DGEMM operation. Thus, the use of this local memory is essential to
achieve high efficiency for DGEMM.

Two cores share one  L1D cache of 2KB. The size of L2D cache is 64KB
and one L2D cache is shared by 16 cores. In the PEZY terminology,
cores that share the L2D cache form a ``city''.  The size of the  L3D cache is
2MB, and 16 ``cities'' share one L3D cache, to form a
``prefecture''. Thus, each prefecture consists of 256 cores, and the
one PEZY-SC chip consists of four prefectures. Finally, the chip is
connected to eight channels of either DDR3 or DDR4 DRAMs.

Thus, the PEZY-SC processor has three levels of cache memories. These
cache memories are {\it not} coherent. Cores which share the same L2D
cache can read the data written by other cores only after explicit
flush operation and barrier synchronization, and the same is true for
the L3D cache and the main memory. Both the flush and barrier
synchronization instructions are provided to each levels of cache. To
be precise, the barrier instruction is available also for one core
(multiple threads).  Clearly, this removal of the cache coherency has
greatly simplified the processor design, and made it possible to
construct a 1024-core processor with three levels of cache.
The line sizes of the L1, L2, and L3 data caches are 64, 256, and 1024
bytes. By changing the line size, the designers of the PEZY-SC
processor kept the bandwidth of L2 and L3 data caches very high. The
bandwidth of L2D cache is the same as that for L1D, and L3D offers
around half of them. 

From the application programmer's point of view, the lack of the cache
coherency does not seem to pose severe limitation, as far as HPC
applications (or their computing kernels) are concerned. For many
applications, programmers know at which moments processors need to
communicate. Moreover, they try to minimize the communication between
cores to achieve high efficiency. Thus, from the point of view of
tuning, the lack of the cache coherency can be regarded as the ability
of the programmer to control the traffic between cores. Moreover, the
barrier synchronization is almost always necessary before the
communication, since otherwise what core A expects that core B has
updated might not be actually updated yet. Efficient
hardware-supported flush and barrier synchronization is thus quite
useful.

The cache for instruction is also multi-level. For I caches, the line
size and the bandwidth are essentially the same for all levels. As far
as all cores run the same and relatively small kernels, this structure
works fine. 

Each PEZY-SC chip has 32 lanes of PCIe (Gen3) interfaces, which is
controlled by integrated two ARM 926 processors. PCIe interfaces
can be used to transfer the data between the main memory of PEZY-SC
processor and the host processor, either by DMA or PIO read/write of
the host processor.

PEZY-SC processor supports a language called PZCL, a dialect of
OpenCL. It supports most of the features of OpenCL, but there are some
limitations in particular when the performance is important (which is
of course almost always the case). The number of software threads
created should be {\it same} as the maximum number of hardware threads
(8192 per chip) to achieve best efficiency. Another difference comes
from the fact that the cache is not coherent. Functions to flush
appropriate levels of cache should be inserted manually to guarantee
the correct result. For small computing kernels, this is not too
difficult, but of course can be a source of hard-to fix bugs.

As one PEZY-SC processor has eight channels of DDR4 DRAMs, the
theoretical peak memory bandwidth is 85GB/s when the DDR4 clock is
1333 MHz. The actual read bandwidth is around 75 GB/s, and STREAM copy
performance is 40 GB/s. The copy performance is low because the write
bandwidth is 1/2 of the read bandwidth.

The read bandwidth of L1, L2 and L3 caches (chip total) are
2000, 2000, and 700 GB/s, respectively. 

\subsection{The ZettaScaler system}
\label{sect:ZettaScaler}

The current generation of the ZettaScaler system (ZettaScaler-1.5)
consists of multiple computing nodes, each of which consists of one
Xeon (E5-v3) processor and four PEZY-SC processors. The Xeon processor
is mounted to a specially-designed  motherboard, and PEZY-SC
processors are mounted to also specially-designed module
boards. The connection between the host Xeon processor and one PEZY-SC
processor is an 8-lane Gen3 PCIe channel. The network between
computing nodes is a standard FDR Infiniband. The largest existing
configuration of ZettaScaler system is a 320-node system called
``Shoubu'', installed at RIKEN ACCC. Smaller systems are installed at
KEK as well as RIKEN AICS.

A very unique feature of the ZettaScaler system is the use of
immersion cooling with fluorocarbon (3M Fluorinert FC-43) coolant.
Compared to previously used oil-based coolant, fluorocarbon coolant
has several advantages like the ease of handling, safety (it is
nonflammable), and smaller coefficient of thermal expansion. The major
disadvantages are the price and potential greenhouse effect, though
the latter is not so severe because of the high vaporization
temperature of the particular coolant actually used.


\section{The overview of HPCG benchmark and implementation on PEZY-SC}
\label{sect:HPCG_on_PEZY_SC}

In this section, we briefly describe the HPCG benchmark itself and
our reference implementation on the PEZY-SC processor.
In subsection \ref{sect:HPCG}, we describe the HPCG benchmark and in
subsection \ref{sect:Implementation_HPCG_on_PEZY_SC}, our implementation of HPCG on PEZY-SC.

\subsection{The HPCG benchmark}
\label{sect:HPCG}

As we've already discussed in section \ref{sect:introduction}, the
HPCG benchmark \cite{Dongarraetal2015, Herouxetal2013} is, according
to its designers, ``designed to measure performance that is
representative of many important scientific calculations, with low
computation-to-data-access ratios.'' As such, it mimics the major
operations of FEM using the CG with  Multigrid solver, on irregular
grid. Unfortunately, the currently available official specification of
HPCG \cite{Herouxetal2013} is rather old, and the algorithm described
there and what is used in the current benchmark code are quite
different. In the following, we first follow \cite{Herouxetal2013} and
then summarize the changes made. 

From the mathematical point of view, the problem solved in HPCG is a
3D diffusion equation discretized using 27-point stencil on a regular
grid of size $(n_x n_{px}, n_y n_{py}, n_z n_{pz})$, where 
$(n_x , n_y , n_z )$ is the size of the grid on each MPI process
and $(n_{px}, n_{py}, n_{pz})$ is the MPI process grid. Thus, the
total number of the MPI processes is $n_{px} n_{py} n_{pz}$.

In the original specification, HPCG solves this problem using the
symmetric Gauss--Seidel preconditioned CG iteration, and the users are
not allowed to change this basic CG algorithm. In particular, the
multigrid method, which is essential if one wants to solve large 3D
problems, is not included.  Thus, not surprisingly, this is changed in
the current specification. Four-stage V-cycle geometric multigrid
preconditioner is used.

What is measured in the HPCG benchmark is the weighted average of the
computing speed of major operations, in particular SymGS, SpMV,
Restriction, Prolongation, DotProduct, and  Waxpby.
Usually, two functions, ComputeSPMV and ComputeSYMGS, dominate the
total computing time and thus determine the performance.

\subsection{Implementation of HPCG on PEZY-SC}
\label{sect:Implementation_HPCG_on_PEZY_SC}

Our reference implementation of HPCG on PEZY-SC is pretty
straightforward. The following six procedures are ported to PEZY-SC
(rewritten using the PZCL language):

\begin{itemize}
\item  SymGS
\item  SpMV
\item  Restriction
\item  Prolongation
\item  DotProduct
\item  Waxpby
\end{itemize}

Both the matrix data and vector data are kept on the memory of
PEZY-SC.  Therefore, only a small amount of data to be transferred for
convergence check and other operations and the boundary data to be
exchanged between nodes are exchanged between the host Xeon processor
and the PEZY-SC processors.

The rewrite using PZCL is pretty straightforward. As noted earlier,
the main point currently we need to care is that the total number of
threads should be actually equal to the available number of hardware
threads. 

Since the
changes which directly take advantage of the regular structure of the
grid are not allowed in the  optimization phase, algebraic block multicolor
ordering\cite{Iwashitaetal2012} is used for the SymGS part.

Table \ref{tab:CGiteration} shows the operations performed in one CG
iteration. Since 4-level V-cycle multigrid method is used, SymGS
routine is called seven times per iteration, and SpMV four times.


\begin{table}[!t]
\renewcommand{\arraystretch}{1.3}
\caption{
    Operations and communication during one CG iteration.
    P and X indicate PEZY-SC and Xeon, $p$ the direction
    vector, and $z$ the preconditioned residual vector, respectively.
}
\label{tab:CGiteration}
\centering
\begin{tabular}{clcccc}
  \hline
  Repeat & Operation & send buf & $p$ & $z$ & cycle\\
  \hline
\multirow{4}{*}{\normalsize 3}
  &FillZero & & & & \multirow{4}{*}{$\searrow$} \\
  &SymGS      & P$\rightarrow$X & & X$\rightarrow$P\\
  &SpMV       & P$\rightarrow$X & & X$\rightarrow$P\\
  &Restriction\\
\hline
  &FillZero\\
\hline
\multirow{2}{*}{\normalsize 3}
  &SymGS & P$\rightarrow$X & & X$\rightarrow$P &
  \multirow{2}{*}{$\nearrow$}\\
  &Prolongation \\
\hline
&SymGS & P$\rightarrow$X & & X$\rightarrow$P &
\multirow{8}{*}{\normalsize CG}\\
& Dotproduct\\
& Waxpby \\
  &SpMV       & P$\rightarrow$X  & X$\rightarrow$P\\
& Dotproduct\\
& Waxpby \\
& Waxpby \\
& Dotproduct\\
\hline
\end{tabular}
\end{table}

\section{HPCG benchmark result}
\label{sect:HPCG_result}

\subsection{Measured Performance}
\label{sect:MeasuredPerformance}

\hyphenation{Ajisai}
We have measured the performance of HPCG on the ``Ajisai''
ZettaScaler system installed in RIKEN AICS. It has the total of 64
Xeon nodes each with four PEZY-SC processors. We made the measurement of
the performance on up to 32 PEZY-SC processors. We assign one PEZY-SC
processor to one MPI process. Therefore, four MPI processes run on
each Xeon processor. The core clock of the PEZY-SC processor is 733
MHz. Memory clock is 1333 MHz. The host CPU is Xeon E5-2618L v3 with
8 cores at 2.3 GHz clock. Each PEZY-SC processor has 32 GB or DDR4
memory, and the host Xeon processor 128 GB.


On 32 PEZY-SC processor, the achieved performance for HPCG 3.0 rating
is
168.06 Gflops (For HPCG 2.4 rating, 189.15 Gflops). The problem size
used is $176^3$ local grid with $4\times 4\times 2$ processor grid,
for the global problem size of $(704, 704, 352)$.

For this particular problem size, HPCG reported convergence after
exactly 50 iterations. 

\begin{figure}[!t]
\centering
\includegraphics[width=2.5in]{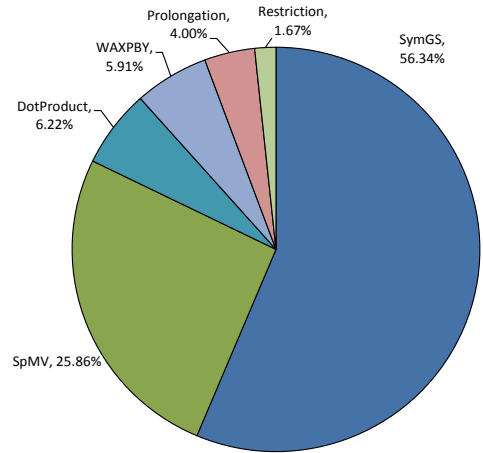}
\caption{Fraction of computing time spent in sections of HPCG
  benchmark code.}
\label{fig:time_pie_chart}
\end{figure}

Figure \ref{fig:time_pie_chart} shows the breakdown of the execution
time. As usual, SpMV and SymGS dominate the execution time. 
The speed of these two sections are 238.4 Gflops and 217.6 Gflops,
respectively. If we calculate the performance per MPI process (or per
PEZY-SC processor), they are 7.45 Gflops and 6.80 Gflops.

For comparison, we measured the single-process performance of the same
code on one PEZY-SC processor. The problem size per MPI process is the
same. The speed of SpMV and SymGS sections are 9.47 Gflops and 8.08
Gflops, respectively. Thus, we can see the overhead of around 20\%,
due to parallel execution (primarily due to communication overhead).

The result of the comparison between single-process and 32-process
performance numbers indicate that the parallelization overhead is,
though not negligible, fairly small, and that single-node performance
is the primary factor that determines the total performance.


\subsection{Performance Analysis of the reference implementation}
\label{sect:PerformanceAnalysis}

In this section, we discuss whether or not the achieved performance of
our reference implementation,
in particular that of the single-chip calculation, is reasonably
optimized or not. We limit our analysis to SpMV, to simplify the
discussion. Since the achieved performance numbers of SpMV and SymGS
are not much different, we believe the analysis of SpMV is
sufficient to discuss the behavior of hardware and software.

The single-chip performance of the SpMV operation on a PEZY-SC
processor is 11.6 Gflops, for the operation at the finest level.
Since the matrix is quite large and each
non-zero element of the matrix is used only once per one SpMV
operation, the performance of SpMV is bandwidth limited. One element
is expressed by one four-byte integer number and one eight-byte
floating-point number, and for this element two floating-point
operations are performed (one multiplication and one addition).
Therefore, to achieve the speed of $x$ Gflops, required memory
read bandwidth is given simply by $(x/2)\times  12  = 6x$ GB/s.
Therefore, 11.6 Gflops means the read performance of 70 GB/s.

The theoretical peak memory bandwidth of a PEZY-SC processor is
85 GB/s, and the actual measured read performance is
around 75 GB/s. Therefore, the peak performance achievable by SpMV is
around 12.5 Gflops. We can see that the performance of 11.6 Gflops is
quite close to what can be achieved.

We can thus conclude that we have successfully ported HPCG on the
PEZY-SC processor, and for the SpMV operation we achieve the
performance very close to the theoretical limit determined by the
throughput of the external memory. Therefore, we now have a good
reference implementation, against which we can measure the effect of
data compression.

\section{Implementation of the SpMV multiplication with data
  compression/decompression  and
  its performance}
\label{sect:SpMV_tuning}

In this section, we investigate  possibilities to reduce the
required memory bandwidth, without changing the basic CG algorithm or 
preconditioner algorithm.
As discussed in section \ref{sect:introduction}, 
one practical possibility to reduce the required memory bandwidth not
explicitly prohibited is the use of on-the-fly data
compression/decompression. The use of the data compression in HPC is
currently an active area of research, and many methods, both lossless
and lossy, have been proposed \cite{JSFI13}.

Many of  the previous proposals of the use of data compression is for
storage and checkpointing, but there are also many studies of the use
of data compression on cache and main memories
\cite{Kaneko2013,Alameldeen2004,Ekman2005}.

So far, the use of data compression on the level of main memory or
cache memory is not quite popular in HPC. In many cases, the
calculation cost of compression/decompression operations is too high.
However, in the case of the access of the sparse matrix in CG
iterations on the PEZY-SC processor, there are several reasons to
expect that the compression technique can be advantageous. First, the
compressed data is reused a number of times. Therefore, the cost of
the compression operation is relatively unimportant. We still need a
fast and efficient decompression algorithm. Second, the PEZY-SC
processor is a fully MIMD and multithreaded processor, which can
generate a very large number of independent memory access
simultaneously. This feature is particularly important, for
table-based data decompression techniques. Each of the 1024 cores of
PEZY-SC processor can execute one load instruction per cycle. On the
other hand, the cores of modern microprocessors with wide-SIMD
execution units can generally issue either one or two load per
cycle. Since the number of cores of these processors are less than 32,
PEZY-SC can be more than one order of magnitude faster in table lookup,
resulting in much better performance in data decompression.

We have tested several implementation of fast
compression/decompression algorithms for simplified implementation of
SpMV operation on the PEZY-SC processor. The original matrix is the
same as what appears in the HPCG benchmark.

So far, the best result is achieved by a simple table-based
compression. In this algorithm, first the entire matrix is scanned and
all unique values in the matrix elements are listed and sorted in the
ascending order. We call this list the value table $V$ and $i$-th
element of $V$ is $v_i$. Therefore, $v_i < v_{i+1}$.  Then, for each
row of the matrix, the non-zero elements are also sorted in the
ascending order. Now we have the list of column indices, sorted by the
actual value of the element. We call this list sorted column list
$S_i$. Now, for each $v_i$ of $V$, we calculate the last position of
that value in the sorted list of non-zero elements, and record that
value to create the list of ``terminal'' indices, $T_i$.

\begin{figure}[!t]

  Original ELL format for a row \\
  \begin{tabular}{|l|ccccc|}
    \hline
    Value & -1 & -1 & 26 & -1 & -1\\
    \hline
    Column & 45 & 49 & 50 & 51 & 65\\
    \hline
  \end{tabular}

  \vskip 3mm
  
  Value table
  
  \begin{tabular}{|l|cc|}
    \hline
    Value & -1 & 26\\
    \hline
  \end{tabular}

  \vskip 3mm

  Compressed expression for a row
  
  \begin{tabular}{|l|ccccc|}
    \hline
    $S_i$  & 45 & 49 & 51 & 65 & 50\\
    \hline
    $T_i$  & 3 & 4 &&&\\
    \hline
  \end{tabular}
  
\caption{The data compression algorithm}
\label{fig:compressionalgorithm}
\end{figure}

The implicit assumption here is that the size of the value table is
small, which is certainly true for HPCG but might not always be true for general
applications of CG method. On the other hand, if one uses
constant-size elements for the area with uniform physical
characteristics, elements with the same values do appear quite often.
Thus, it is quite likely that our compression method would work for
the large fraction of the rows of the original matrix, even if we limit
the size of the value table to be small. In the case of the matrix in
the HPCG SpMV operation, the size of $S_i$ per one node is 27,
and $T_i$ can be two. Thus, one row of the matrix is now compressed to
$(27+2)\cdot 4 = 116$ bytes, from the original size of 324 bytes.

We can further compress the list of $S_i$ in the following way. First,
we convert the actual values of column indices to the value relative
to the diagonal element. Then, we register the pattern of this
relative displacement of column indices to a table to perform the data
compression. Since the majority of the nodes have the same index
pattern for relative displacement, this compression can effectively
reduce the size of the index array to $4N$ bytes, where $N$ is the
matrix dimension. Thus, instead of 116 bytes per node, we have now 12
bytes per node. If we register to this table the values of matrix
elements themselves, we can probably reduce the size by another factor
of three, resulting in essentially four bytes per node.

We can probably further compress the data by applying a simple
run length compression to the final table.

\begin{table}[!t]
\renewcommand{\arraystretch}{1.3}
\caption{The effect of the data compression algorithms applied to SpMV
  operation}
\label{tab:compression}
\centering
\begin{tabular}{lcc}
  \hline
 Compression method & Measured performance & Theoretical performance \\  
  \hline
  Original  & 11.6GF & 12.5GF\\
  Data Table  & 15.9GF & 34.8GF\\
  Data+Index Table  & 32.4GF & 326GF\\
\hline
\end{tabular}
\end{table}

So far, We have actually implemented the first two compression
schemes. Table \ref{tab:compression} show the resulting performance.
The first approach of compressing the data array only should
theoretically give around a factor of three speedup, and actually
realized the speedup by 50\%. The second one, in which both the index
array and the data array are compressed, should theoretically give
around a factor of 25 speedup, and actually achieved the speedup by a
factor of 2.8.

The reason why the actual speedup is much smaller than the theoretical
limit is simply that to estimate the theoretical limit we ignore the
access cost of the input vector, which is currently accessed
indirectly with rather large address offsets in the innermost loop.
With reordering of the vector and matrix we might be able to
improve the performance further.

List \ref{code:samplecode} shows the conceptual code for data and
index compression. The fact  indices and data are compressed means
that their actual values are obtained by table lookup
operations. Thus, on modern
microprocessors with wide SIMD instruction sets, it would be difficult
to achieve reasonable performance with the compression algorithm,
since the throughput of indirect access operations are generally low.
The fully-MIMD, non-SIMD nature of the PEZY-SC processor is critical
to achieve the actual speedup.

\begin{lstlisting}[label=code:samplecode,caption=A sample code with both data and index arrays compressed]

for(int i = 0; i < n; i++){
    y[i] = 0;
    const int type = columnDiffType[i];
    int idx = 0;
    for(int valueIdx = 0; valueIdx < valueCount;
        valueIdx++){
        const double a_ij = value[valueIdx];
        for(;idx < valueIdxEnd[i][valueIdx]; idx++){
            const int j = i + columnDiff[type][idx];
            const double x_j = x[j];
            y[i] += a_ij * x_j;
        }
    }
}
\end{lstlisting}

Note, however, that the problem here is the number of independent
memory access per cycle, and not the difference between SIMD and MIMD
architecture. The gather/scatter functions of modern SIMD
microprocessors are clearly still in their infancy, and might be
improved in the future.


\section{Summary}
\label{sect:discussion}

In this paper, we report the effect of data compression/decompression
algorithms for the SpMV multiplication on  the ZettaScaler system, in
which the 1024-core, MIMD ultra-many-core PEZY-SC processors are used
as  accelerators. We have used the matrix generated by HPCG benchmark
code as the example. 
Usually, 
the performance of the well-optimized implementation of HPCG is
limited by the bandwidth of the sequential read access of the external
memory of the processor (or accelerator). In the case of a PEZY-SC
processor, the theoretical limit of the read bandwidth is 85 GB/s,
and actual measured bandwidth is 75 GB/s. Thus, the performance of
SpMV and SymGS operations are limited to around 10 Gflops. The actual
performance achieved is close to this number.

The theoretical speedup by data and index compression is as large as a
factor of 25. We actually achieved the speedup of a factor of
2.8 for the SpMV operation. Even though the actual achieved improvement
is much smaller than the theoretical maximum, we have demonstrated that
the used of data compression/decompression can actually improve the
performance of SpMV multiplication on the PEZY-SC processor. We
therefore conclude that the use of data compression/decompression
will be quite useful technique to improve the performance of SpMV
operations in FEM applications on the current and future
high-performance processors.

\section*{Acknowledgment}

\ifnum \nonbdmode=1

The authors would like to thank people in PEZY Computing/ExaScaler for
their invaluable help in solving many problems we encountered while porting and
tuning HPCG.  Part of the research covered in this paper research was
funded by MEXT's program for the Development and Improvement for the
Next Generation Ultra High-Speed Computer System, under its Subsidies
for Operating the Specific Advanced Large Research Facilities.

\fi





\newcommand{\noopsort}[1]{} \newcommand{\printfirst}[2]{#1}
  \newcommand{\singleletter}[1]{#1} \newcommand{\switchargs}[2]{#2#1}


\end{document}